\documentclass[aps,prd,preprint,groupedaddress,nofootinbib]{revtex4}
\pdfoutput=1
\usepackage{amsmath,amssymb,amsbsy,color}
\usepackage{latexsym}
\usepackage{graphics}
\usepackage{graphicx}
\usepackage{psfrag,bbm}

\newcommand{\be}{\begin{equation}}
\newcommand{\ee}{\end{equation}}
\newcommand{\bea}{\begin{eqnarray}}
\newcommand{\eea}{\end{eqnarray}}
\newcommand{\ba}{\begin{array}}
\newcommand{\ea}{\end{array}}

\newcommand{\psibar}{\overline{\psi}}
\newcommand{\chibar}{\overline{\chi}}

\def\bqry{\begin{eqnarray}}
\def\eqry{\end{eqnarray}}

\newcommand{\tr}[1]{{\rm Tr}(#1)}
\newcommand{\str}[1]{{\rm Str}(#1)}
\newcommand{\sdet}[1]{{\rm Sdet}(#1)}
\newcommand{\Det}[1]{{\rm Det}(#1)}

\begin{document}

\preprint{CERN-PH-TH-2010-308}
\preprint{HU-EP-11/02}
\preprint{IFT-UAM/CSIC-11-01}
\title{Spectral density of the Hermitean Wilson Dirac operator: a NLO computation in chiral perturbation theory}

\author{Silvia Necco$^1$ and Andrea Shindler$^{2,3}$\footnote{Heisenberg Fellow}}

\affiliation{
$^1$CERN, Physics Departement, 1211 Geneva 23, Switzerland\\
$^2$Instituto de F\'{\i}sica Te\'orica UAM/CSIC\\
Universidad Aut\'onoma de Madrid, Cantoblanco E-28049 Madrid, Spain \\
$^3$Institut f\"ur Physik, Humboldt Universit\"at zu Berlin, Newtonstrasse
15, 12489 Berlin, Germany\\
}

%
\begin{abstract}
We compute the lattice spacing corrections to the spectral density of
the Hermitean Wilson Dirac operator using Wilson Chiral Perturbation
Theory at NLO. We consider a regime where the quark mass $m$ and the lattice
spacing $a$ obey the relative power counting $m\sim a\Lambda_{\rm
  QCD}^2$: in this situation discretisation effects can be treated as
perturbation of the continuum behaviour. While this framework fails
to describe lattice spectral density close to the threshold, it allows
nevertheless to investigate important properties of the spectrum of
the Wilson Dirac operator. We discuss the range of validity of our results
and the possible implications in understanding the phase diagram of Wilson fermions.

\end{abstract}

\maketitle

\section{Introduction}
\label{sec:intro}

Simulations with Wilson fermions at light quark masses are nowadays feasible thanks to 
considerable algorithmic 
improvements~\cite{Hasenbusch:2002ai,Luscher:2005rx,Urbach:2005ji,Clark:2006fx,Luscher:2007se} 
developed in the last years.
Since chiral symmetry is explicitly broken, the spectrum of the Wilson Dirac operator is not protected from 
arbitrarily small eigenvalues, which might induce instabilities in numerical 
simulations~\cite{DelDebbio:2005qa}. It is therefore very important to have a theoretical understanding of the 
properties of the low-end spectrum 
of the Wilson Dirac operator. Moreover, spectral observables can be efficiently used to extract relevant
quantities such as the quark condensate, like recently implemented in \cite{Giusti:2008vb}, yielding a 
further strong motivation to investigate the impact of lattice artefacts on the eigenvalues spectrum.

In the continuum and in an infinite volume, the (renormalised) spectral density of the massive Dirac operator has a 
threshold given by the (renormalised) quark mass. Lattice artefacts are expected to change both the location 
of the threshold and the shape of the spectral density close to the threshold. Moreover, in a finite box of 
volume $V$ one expects that when $\gamma\Sigma V\simeq 1$, where $\gamma$ is an eigenvalue of the massless 
Dirac operator and $\Sigma$ the chiral condensate, finite-size effects become important and can also induce 
relevant deformations of the spectral 
density with respect to the infinite-volume case.

From the numerical point of view, some information about the low-end spectrum is provided by lattice 
simulations with Wilson fermions carried out in the past years. For $N_f=2$ unimproved Wilson fermions, 
empirical observations~\cite{DelDebbio:2005qa} indicate that the 
median of the spectral gap distribution is linearly proportional to the quark mass $m$, while the width 
is basically independent on $m$ and scales like $\sim a/\sqrt{V}$, where $a$ is the lattice spacing. 
On the other hand, for the $O(a)$-improved theory, the situation is less clear \cite{DelDebbio:2007pz} 
and those properties have not been confirmed. In \cite{Giusti:2008vb} the mode number of the $O(a)$-improved 
Wilson Dirac operator is computed, finding a nearly linear behavior up to $\simeq 100$ MeV above the threshold. 
At the low end of the spectrum, a significant deviation from the continuum expectation is observed.

From the theoretical side, Wilson chiral perturbation theory~\cite{Sharpe:1998xm,Rupak:2002sm} (W$\chi$PT) is the tool which 
provides a systematic description of low-energy properties of lattice Wilson QCD including the leading 
discretisation effects. When approaching the chiral limit at finite lattice spacing, one enters in the 
regime where $m\sim a^2\Lambda_{\rm QCD}^3$, which is where discretisation effects compete with the quark 
mass to the explicit breaking of chiral symmetry. Lattice artefacts induce a non-trivial phase diagram, 
and two different scenario have been foreseen: in the so-called Aoki scenario there is a range of quark 
masses (in the Aoki phase~\cite{Aoki:1983qi}) where there are two massless pions.
On the other hand, in the so-called Sharpe-Singleton scenario~\cite{Sharpe:1998xm} 
there is a first-order phase transition, and the three pions remain massive in the chiral limit. 
From the point of view of W$\chi$PT, in this regime the 
discretisation effects appear already at leading order in the chiral expansion. 
Lattice artefacts in the infinite volume spectrum of the Hermitean Wilson Dirac 
operator have been computed in this regime in \cite{Sharpe:2006ia}, although a working 
framework has been only found by assuming additional conditions on the couplings of W$\chi$PT 
associated to discretization effects.

In a recent study \cite{Damgaard:2010cz,Akemann:2010em} the same power counting for $a$ has been adopted, but in a 
finite-box in the $\epsilon$-regime\footnote{Standard W$\chi$PT has been extended to the $\epsilon$-regime 
in~\cite{Shindler:2009ri,Bar:2008th,Bar:2010zj}.}.
In this case the LO predictions of W$\chi$PT can be obtained also by means of a ``modified'' Random 
Matrix Theory which includes $O(a^2)$ effects. The main difficulty of this computation is that it 
involves exact integrals over the zero modes which must be defined at fixed topological charge introduced via the number of real modes of the Dirac operator, which nevertheless has an intrinsic ambiguity for Wilson fermions at finite lattice spacing.

In this work we will study the discretisation effects in the spectrum 
of the Dirac operator in a different regime. We consider W$\chi$PT with the 
power counting $m\sim a\Lambda_{\rm QCD}^2$ (GSM regime), where lattice artefacts appear only at 
next-to-leading order in the chiral effective theory, and can hence be treated as a perturbation. 
This simplifies considerably the computation; it allows nevertheless to extract important information 
about the impact of lattice artefacts in the spectral density. 

This paper is structured as follows: in sec.~\ref{sec:qcd} we recall definitions and properties of the 
spectral density of the Wilson Dirac operator; in sec.~\ref{sec:graded} we present the setup for 
W$\chi$PT needed for the computation; in sec.~\ref{sec:density} we give the details about the calculation 
of the spectral density, including the finite-volume corrections (in the $p$-regime); in sec.~\ref{sec:disc} 
we discuss our results and show a comparison with numerical data. All technical details about the computation are deferred in the Appendices.

\section{Spectral density for the Wilson Dirac operator}
\label{sec:qcd}

We start considering QCD for $N_f=2$ degenerate quarks in infinite volume
with current quark mass $m$.
The average spectral density of the massless Hermitian Dirac operator $-iD$ can be defined as
\be
\rho_D(\gamma,m) \equiv \lim_{V \rightarrow \infty}
\frac{1}{V}\sum_k \langle \delta \left(\gamma - \gamma_k \right) \rangle,
\ee
where $\gamma_k$ are the eigenvalues of the massless Hermitian Dirac operator and $\langle \ldots \rangle$ 
indicates the usual path-integral average.
The Banks-Casher relation~\cite{Banks:1979yr} tells us that the spectral density
is related to the chiral condensate $\Sigma$ in the following way
\be
\rho_D\left(\gamma,m\right) = \frac{\Sigma}{\pi}\left[1+O(\{|\gamma|,m\}/\Lambda_{QCD}) \right] .
\label{eq:BC}
\ee
One useful way to determine this relation in continuum 
QCD~\cite{Chandrasekharan:1994cq,Verbaarschot:1995yi} consists in adding a 
valence quark $\psi_v$ of mass $m_v$ to the theory. Using the spectral decomposition for the
valence chiral condensate one obtains
\be
\langle \psibar_v \psi_v \rangle = - \int d\gamma \frac{\rho_D(\gamma,m)}{i \gamma + m_v} .
\label{eq:cond_spect}
\ee
This relation can be inverted because the spectral density is independent on the valence 
quark mass
\be
{\rm Disc}\left[\langle \psibar_v \psi_v \rangle \right]{\large|}_{m_v = -i \gamma} = -2 \pi \rho_D(\gamma,m),
\label{eq:disc}
\ee
where ${\rm Disc}$ indicates the discontinuity across the imaginary valence quark mass.
Strictly speaking eq.~\eqref{eq:cond_spect} is ultraviolet divergent also after renormalisation 
of the quark masses 
and the gauge coupling constant and one must introduce a cutoff
in the integration range. The ultraviolet divergences turn out to cancel when computing the
discontinuity of the valence scalar condensate in eq.~\eqref{eq:disc}. 
Thus it is enough to compute the valence condensate for real masses and analytically 
continuing the resulting expression for complex masses.
Eq.~\eqref{eq:disc} can be used naturally in chiral perturbation theory and 
one obtains~\cite{Verbaarschot:1995yi} the Banks-Casher relation~\eqref{eq:BC} and 
the NLO corrections~\cite{Osborn:1998qb,Giusti:2008vb}.

We now discretise our $N_f=2$ continuum QCD action on a lattice of spacing $a$ and we 
consider Wilson fermions. All our considerations can be generalised to the case 
of Wilson twisted mass. 
To study the spectral density of the Wilson operator one has to take into account that
the Wilson operator $D_W$ is neither Hermitian nor anti-Hermitian and it has complex eigenvalues.
It is thus advantageous to define the Hermitian Wilson-Dirac operator $Q_m$ 
\be
Q_m = \gamma_5 \left(D_W +m \right)=Q_m^{\dagger}.
\ee
We indicate now the spectral density of $Q_m$ as $\rho_Q(\lambda,m)$ and the one of 
$Q_m^2$ as $\rho(\alpha,m)$, where $\lambda$ and $\alpha$ label the eigenvalues of the 
corresponding operators.

If the continuum theory has been regulated on a lattice the ultraviolet divergences 
of the chiral condensate appear as power law and logarithmic divergences in the lattice spacing $a$.
Moreover, when using Wilson fermions the lack of chiral symmetry
implies that the scalar condensate is not directly related to the spectral density as in the continuum.
It is thus not obvious how to work out
the renormalisation and O($a$) improvement of the spectral density starting from 
eq.~\eqref{eq:disc}
or from the corresponding version for Wilson lattice QCD.
To study the renormalisation and the O($a$) improvement
of the spectral density one needs to relate it to correlation functions
of local operators where standard arguments on renormalisability and O($a$) improvement can be 
applied.

It has been shown in ref.~\cite{Giusti:2008vb}, 
using chain correlators of scalar and pseudoscalar densities how the spectral density of $Q_m^2$ 
renormalises. In the following we will be mostly interested
in the spectral density of $Q_m$ that renormalises as follows
\be
\left[\rho_Q\right]_R(\lambda, m_R)= Z_P\rho_Q(Z_P \lambda,m),
\ee
where $m_R$ is the renormalised quark mass.
Moreover, additionally to the standard improvement of the action and local operators 
there are O($am$) cutoff effects which need to be removed to fully improve the spectral 
density~\cite{Giusti:2008vb}. 
This analysis guarantees that with Wilson fermions the spectral density is a well defined renormalisable 
quantity and with a well defined Symanzik expansion. We can thus compute the spectral density
in the chiral effective theory using a generalisation of eq.~\eqref{eq:disc} 
for Wilson chiral perturbation theory.\\
A further spectral observable which can be defined and measured in a lattice simulation is the 
integrated spectral density
\begin{equation}\label{eq:modenum}
N(\Lambda_1,\Lambda_2,m)= \int_{\Lambda_1}^{\Lambda_2} d\lambda~ \rho_Q(\lambda,m),\;\;\;\; \Lambda_2\geq \Lambda_1\geq m,
\end{equation}
which represents the density of modes in the interval between $\Lambda_1$ and $\Lambda_2$;
it contains the same physical information as the spectral density and 
satisfies $N_R(\Lambda_{1,R},\Lambda_{2,R},m_R)=N(\Lambda_1,\Lambda_2,m)$, i.e. it is a 
renormalisation-group invariant \cite{Giusti:2008vb}.

To relate a partially quenched condensate with the spectral density of $Q_m$
it is convenient to introduce a doublet of degenerate twisted mass fermions $\chi_v$ 
with a mass term $i \mu_v \gamma_5 \tau^3$, where $\tau^3$ is the third Pauli matrix. 
The relation between the condensate and the spectral density is now different
but can be worked out and it reads~\cite{Sharpe:2006ia} 
\be
\langle \chibar_v \gamma_5 \tau^3 \chi_v \rangle = \int_{-\infty}^\infty d\lambda 
\frac{\rho_Q(\lambda,m)+\rho_Q(-\lambda,m)}{\lambda + i \mu_v} .
\label{eq:pqcond}
\ee
The need to symmetrise in $\lambda$ the spectral density in the numerator of~\eqref{eq:pqcond}
can be understood from the lack of chiral symmetry of the Wilson operator which renders the spectral
density not symmetric when $\lambda \rightarrow -\lambda$. Inverting eq.~\eqref{eq:pqcond}
one obtains
\be
{\rm Disc}\left[\langle \chibar_v \gamma_5 \tau^3\chi_v \rangle \right]|_{\mu_v = i \lambda} = 2 i \pi 
\left[\rho_Q(\lambda,m)+\rho_Q(-\lambda,m)\right] \,
\label{eq:disc_latt}
\ee
where the untwisted mass in the valence sector coincides with the sea quark mass $m$.
Eq.~\eqref{eq:disc_latt} is our starting point for the computation of the spectral
density in Wilson chiral perturbation theory. 

We stress that in chiral perturbation theory the problem of power law divergences is absent 
because those are mapped to the presence of the so called high-energy constants
usually indicated with $H$.
As we will see in sect.~\ref{sec:density} this is indeed the case for the
expectation value of the pseudoscalar condensate in the chiral effective theory.
This implies that in the effective theory the computation of the discontinuity is a well defined procedure.
It remains to make sure that the spectral density computed in the effective theory through the discontinuity
of the partial quenched chiral condensate does not overlook additional cutoff effects 
which go beyond the standard O($a$) corrections stemming from the action or the local operators.
Our previous discussion and the results of ref.~\cite{Giusti:2008vb} guarantees
that the only additional cutoff effects are of O($am$) and as we will see in sect.~\ref{sec:graded}
these cutoff effects are of subleading order in our power counting of Wilson chiral perturbation theory.

\section{Spectral density in Wilson Chiral Perturbation Theory    }
\label{sec:graded}
In order to investigate the discretisation effects in the spectral density of the Hermitian 
Wilson Dirac operator we compute the 
valence pseudoscalar density in the framework of Partially Quenched Wilson chiral perturbation 
theory (PQW$\chi$PT).
According to eq.~\eqref{eq:pqcond}, we consider $N_f=2$ sea quarks with bare mass $m$ and a 
doublet of valence twisted mass quarks with $(m+i\mu_v\tau^3)$.
We formulate the effective theory on a finite volume $V=L^3 T$ and we keep the quark masses 
in the $p$-regime, corresponding to the power counting
\be
m,\mu_v\sim O(p^2),\;\;\;1/L,1/T \sim  O(p)
\label{eq:power_counting}
\ee
in terms of the momenta $p$.
The extension of the effective theory to the case of non-zero lattice spacing is done in 
two steps: after matching the lattice QCD action with the appropriate Symanzik continuum 
effective action \cite{Symanzik:1983dc,Symanzik:1983gh}, one writes down a chiral Lagrangian 
which contains the standard continuum terms plus additional operators that transform under 
chiral symmetry as the operators of the Symanzik effective theory. For the Wilson action, 
this has been studied in \cite{Sharpe:1998xm,Rupak:2002sm}.
When introducing lattice artefacts in the chiral effective theory, we have to define as 
usual the relative power counting between the quark mass and the lattice spacing $a$. 
In this work we adopt a counting corresponding to the so-called 
GSM (\emph{Generically Small quark Mass}) regime  \cite{Sharpe:2004ps,Sharpe:2004ny},
where $m,\mu_v \sim a\Lambda_{\rm QCD}^2$: in this case the explicit breaking of chiral 
symmetry is dominated by the quark mass, and lattice artefacts can be treated as perturbations.

Partially quenching can be implemented in the Chiral Effective theory by means of two 
techniques, namely the graded-symmetry  method \cite{Bernard:1992mk,Bernard:1993sv} and 
the replica method \cite{Damgaard:2000gh}. In the first one, one introduces ``ghost'' 
quarks, whose determinant cancels the one of the valence quarks. In the second one, 
one enlarges the valence sector to $N_r$ flavors and one eliminates the corresponding 
determinant by sending $N_r\rightarrow 0$. The equivalence of the two methods has been 
shown at the perturbative level in \cite{Damgaard:2000gh}.  
In this section we will focus on the graded-symmetry method; we checked nevertheless 
that the same results can be obtained with the replica method, which we summarise in 
App. \ref{sec:app_b}. For our specific case we have to consider a chiral Effective 
Theory with a graded symmetry group SU$(4|2)_L$$\times$SU$(4|2)_R$ spontaneously 
broken to SU$(4|2)_{R+L}$. 
The formulation we adopt for our computation using the SU$(4|2)$ effective theory
is based on an extension of the framework proposed in~\cite{Giusti:pv} for the SU$(3|1)$ case.
We consider a mass matrix of the form
\begin{equation}\label{eq:mass}
\mathcal{M}={\rm diag} (\underbrace{m}_{2\times 2\; {\rm sea}}, \underbrace{m+i\mu_v\tau^3}_{2\times 2\; {\rm val}},\underbrace{m+i\mu_v\tau^3}_{2\times 2\; {\rm ghost}}).
\end{equation}
The quark mass in the ghost sector has been already set equal to the mass in the valence sector. Moreover, the untwisted part of the valence quark mass has to be equal to the sea quark mass.
We parametrise the pseudo Nambu-Goldstone bosons by the field $U\in$ SU$(4|2)$
\begin{equation}
U(x)=u_Ve^{2i\xi(x)/F}u_V,\;\;\;\;\xi=\sum_a \xi^a T^a,
\end{equation}
where $F$ is as usual the pseudoscalar decay constant, and $T^a,a=1,\ldots,35$ represent the generators of the corresponding Lie algebra, which satisfy 
\begin{equation}
\str{T^aT^b}=\frac{g^{ab}}{2}.
\end{equation}
 We refer to App. \ref{sec:app_a} for the explicit form of $g^{ab}$ (eq.~\eqref{eq:gab}), and for a summary of conventions and properties of the SU$(m|n)$ group. 
The constant field $u_V$ represents the ground state of the theory, which can be obtained by minimising the potential in the LO Chiral Lagrangian. 
The solution of the potential minimisation yields \footnote{In the graded symmetry formulation of PQ$\chi$PT there are subtleties concerning the minimisation of the potential. We will discuss them in more detail in sec. \protect\ref{sec:density}.} 
\begin{equation}\label{eq:grstate}
u_V={\rm diag} (\underbrace{1}_{2\times 2\; {\rm sea}}, \underbrace{e^{i\tau^3\omega_0/2}}_{2\times 2\; {\rm val}},\underbrace{e^{i\tau^3\omega_0/2}}_{2\times 2\; {\rm ghost}}),
\end{equation} 
where 
\begin{equation}\label{eq:omega_mp}
\sin\omega_0=\frac{\mu_v}{m_P},\;\;\;\cos\omega_0=\frac{m}{m_P},\;\;\;\;m_P=\sqrt{m^2+\mu_v^2}.
\end{equation}
In the language of twisted mass Wilson theory, $m_P$ is the so-called \emph{polar mass}. 

Taking into account our power counting, the full NLO Chiral Lagrangian in PQW$\chi$PT can be written as
\begin{equation}\label{eq:fullL}
\mathcal{L}=\mathcal{L}_{2}+\mathcal{L}_{4}+\mathcal{L}_a.
\end{equation}
In the GSM regime, the LO Lagrangian can be written in the same form as the continuum one
\begin{eqnarray}
\mathcal{L}_{2} & = & \frac{F^2}{4}\left\{
\str{\partial_\mu U\partial_\mu U^\dagger}-2B\str{\mathcal{M}U^\dagger+ U\mathcal{M}^\dagger}\right\}, 
\end{eqnarray}
provided we substitute the quark mass with the so-called \emph{shifted} mass \cite{Sharpe:1998xm}, which incorporates the leading $O(a)$ corrections. In the following we assume that $m$ in eq.~\eqref{eq:mass} includes already this shift. The coupling $B$ is related to the chiral condensate $\Sigma$, $B=\Sigma/F^2$.  \\    
From $\mathcal{L}_{2}$ we can extract the propagator
\begin{equation}
\langle \xi^a(x)\xi^b(y)\rangle =g^{ab}G_V^1(x-y;M_a^2)+2B(m_P-m)h^{ab}G_V^2(x-y;M_a^2),\;\;\;\;(a,b=1,\dots,35),
\end{equation}
with
\begin{equation}
G_V^r(x,M^2)\equiv\frac{1}{V} \sum_p \frac{e^{ipx}}{(p^2+M^2)^r},\;\;\;(r\geq 1),\;\;\;p=2\pi\left(\frac{ n_1}{L},\frac{ n_2}{L},\frac{ n_3}{L},\frac{n_4}{T}\right),\;\;\; n_{1,2,3,4}\in \mathbb{Z}.
\end{equation}
We recall that \cite{Hasenfratz:1989pk}
\begin{equation}\label{eq:fsprop}
G_V^r(0,M^2)=G^r(0,M^2)+g_r(M^2),
\end{equation}
where
\begin{equation}\label{eq:ginf}
G^r(x,M^2)=\frac{1}{(2\pi)^4}\int d^4p  \frac{e^{ipx}}{(p^2+M^2)^r},
\end{equation}
is the infinite-volume contribution, while $g_r(M^2)$ represents the finite-volume correction and is UV-finite. A specific representation of $g_r(M^2)$  will be considered in App. \ref{sec:app_c}. 
The matrix $h^{ab}$ is defined in App. \ref{sec:app_a}, eq.~\eqref{eq:hab}. The mass term $M_a^2$ appearing in the propagator is given by
\begin{equation}
M_a^2=\left\{  \begin{array}{lll} 
M^2_{ss} = & 2mB, & a=1,\ldots,3\\
M^2_{sv} = & (m+m_P)B, & a=4,\ldots,11,16,\ldots,23\\
M^2_{vv} = & 2m_PB, & a=12,\ldots,15,24,\ldots,35. 
\end{array}
\right.
\end{equation}
The continuum NLO PQ Lagrangian reads \cite{Gasser:1984gg}
\begin{eqnarray}
\mathcal{L}_{4} & = & 2 B L_4\str{\partial_\mu U^\dagger\partial_\mu U }\str{\mathcal{M}U^\dagger+\mathcal{M}^\dagger U}
+2BL_5\str{\partial_\mu U^\dagger\partial_\mu U ( \mathcal{M}U^\dagger+\mathcal{M}^\dagger U)  }   \nonumber\\
& & -4B^2L_6\str{U^\dagger\mathcal{M}+\mathcal{M}^\dagger U}\str{U^\dagger\mathcal{M}+\mathcal{M}^\dagger U} \\
& & -4B^2L_7\str{\mathcal{M}^\dagger U-\mathcal{M}U^\dagger}\str{\mathcal{M}^\dagger U-\mathcal{M}U^\dagger}\nonumber\\
& & -4B^2L_8\str{\mathcal{M}U^\dagger\mathcal{M} U^\dagger+ \mathcal{M}^\dagger U\mathcal{M}^\dagger  U    } -4B^2H_2\str{\mathcal{M}^\dagger \mathcal{M}},\nonumber\end{eqnarray}
where we have written down only terms which can enter in our specific computation.
The PQ Chiral Lagrangian encoding discretisation effects up to $O(a^2)$ can be written as \cite{Rupak:2002sm,Bar:2003mh,Aoki:2003yv}
\begin{eqnarray}
\mathcal{L}_a & = & -2B\hat{a}W_6\str{\mathcal{M}U^\dagger+\mathcal{M}^\dagger U}\str{U+U^\dagger}
- 2B\hat{a}W_7\str{\mathcal{M}^\dagger U-\mathcal{M}U^\dagger}\str{U-U^\dagger}\nonumber \\ 
&&- 2B\hat{a}W_8\str{\mathcal{M}U^{\dagger 2}+\mathcal{M}^\dagger U^2}
-\hat{a}^2W_6'(\str{U+U^\dagger})^2-\hat{a}^2W_7'(\str{U-U^\dagger})^2 \nonumber\\
& &- \hat{a}^2W_8'\str{U^2+U^{\dagger 2}},
\end{eqnarray}
where $\hat{a}=2W_0a$ has dimension [energy]$^2$ and $W_0$ is the LO low-energy coupling absorbed in the shifted mass $m$. Also in this case we have disregarded terms which are not relevant for our computation.

\section{Calculation of the spectral density with SU(4$|$2) graded group method}\label{sec:density}
In this section we compute the spectral density at NLO in W$\chi$PT, making use of the definitions and assumptions introduced in the previous section.
The observable from which we start is the partially quenched twisted pseudoscalar condensate made of valence quarks defined in eq.~\eqref{eq:pqcond}. In order to extract it we have to introduce source terms in the Chiral Lagrangian by means of
\begin{equation}
\mathcal{M}\rightarrow \mathcal{M}_{p_3}=\mathcal{M}+p_3\hat{\tau}^3, \;\;\;\mathcal{M}^\dagger\rightarrow \mathcal{M}^\dagger_{p_3}=\mathcal{M}^\dagger-p_3\hat{\tau}^3,
\end{equation} 
where $\mathcal{M}$ is the mass matrix introduced in eq.~\eqref{eq:mass}, and $\hat{\tau}^3$ has non-zero components only in the valence sector 
\begin{equation}
\hat{\tau}^3={\rm diag} (\underbrace{0}_{2 \times 2\; {\rm sea}},\underbrace{\tau^3}_{2\times 2\; {\rm val}},\underbrace{0}_{2\times 2\; {\rm ghost}}).
\end{equation}
The pseudoscalar density is obtained by deriving the action associated to the Chiral Lagrangian in eq.~\eqref{eq:fullL} with respect to $p_3$ 
\begin{equation}
\mathcal{P}^3(x)=  \frac{\delta}{\delta p_3(x)}\mathcal{S}|_{p_3=0}.
\end{equation}
At LO ($O(p^0)$) we obtain the (continuum) expectation value
\begin{equation}\label{eq:p3lo}
\langle\mathcal{P}^3\rangle_{LO}=2i\Sigma\sin\omega_0.
\end{equation}
The NLO result is 
\begin{eqnarray}\label{eq:p3nlograd}
\langle\mathcal{P}^3\rangle_{NLO} &= & 2i\Sigma\sin\omega_0\Bigg\{1+\delta\cot\omega_0 + \frac{1}{F^2}\Bigg[\frac{1}{2}G_V^1(0,M^2_{vv})-2G_V^1(0,M^2_{sv})+32L_6M^2_{ss}\\
&-&\frac{1}{2}(M^2_{vv}-M^2_{ss})G_V^2(0,M^2_{vv})+4M^2_{vv}(H_2+2L_8) 
+ 8\hat{a}\left(2W_6+W_8\cos\omega_0\right)\Bigg]\Bigg\} \nonumber, 
\end{eqnarray}
where $\delta$ is an $O(p^2)$ quantity that represents the correction to the ground state angle $\omega_0$ due to lattice artefacts. In the following subsection we discuss the computation of this effect.

\subsection{NLO correction to the vacuum}\label{sec:vac}
The $O(p^2)$ terms in the chiral Lagrangian give rise to a shift in the vacuum angle $\omega_0$ \cite{Sharpe:2004ny}, which must be computed by minimising the NLO potential. An important fact is that at this order the continuum $L_i$ do not contribute to this realignment, and only the discretisation effects must be taken into account. 
Since this is an important point of the calculation, we start by recalling how the continuum ground state given in eq.~\eqref{eq:grstate} is obtained. 
Using the equations of motions one can show that $u_V$ has to commute with the matrix $\mathcal{M}^\dagger\mathcal{M}$, which implies that $u_V$ is diagonal. Therefore we can parametrise it as
\begin{equation}\label{eq:vac}
u_V={\rm diag} \left(e^{i\frac{\phi_1}{2}},e^{i\frac{\phi_2}{2}}, e^{i\frac{\phi_3}{2}},e^{i\frac{\phi_4}{2}},e^{i\frac{\phi_5}{2}}, e^{i\frac{\phi_6}{2}}\right),
\end{equation}
where $\pi < \phi_1, \ldots , \phi_4 \le \pi$ are standard phase factors.
The condition $\sdet{u_V}=1$ requires
\begin{equation}
\phi_1+\phi_2+\phi_3+\phi_4-\phi_5-\phi_6=0.
\end{equation}
As suggested in~\cite{Sharpe:2001fh}, we perform an analytic continuation for the ghost components
\begin{equation}
\phi_5=i\hat{\phi_5},\;\;\;\phi_6=i\hat{\phi_6},
\end{equation}
where now $\hat{\phi}_{5,6}$ are real variables taking values along the real axis.
The LO potential can be separated in quark and ghost components
\begin{equation}
V_{LO} =  V_{q,LO}+V_{g,LO},
\end{equation}
with
\begin{eqnarray}
V_{q,LO} & =&  2 [{m_P} (\cos (\omega_0 -{\phi_3})+\cos (\omega_0 +{\phi_4}))+{m} (\cos {\phi_1}+\cos {\phi_2})],\\
V_{g,LO} & = & -2m_P [\cos (\omega_0 -i{\hat{\phi}_5})+\cos (\omega_0   +i{\hat{\phi}_6})].
\end{eqnarray}
The potential is complex, and a minimisation procedure in the usual sense is not applicable. The minimisation of the potential
in Euclidean field theory is equivalent to perform a saddle-point expansion of the functional integral, and the saddle-point expansion can be
performed even if the potential is complex. Our prescription, following~\cite{Golterman:2005ie}, is to find the saddles
for complex $\hat{\phi}_{5,6}$ and then deform the contour of integration for each field variable in order to pass through the saddle points.
The saddle points are chosen following two criteria: a) in order to maximise the value of the real part of the potential at the saddle;
b) in order to have the direction where the real part of the potential rises more steeply consistent with the possibility
of deforming the contour integration without encountering the other saddle points.
This procedure has been discussed in detail in~\cite{Golterman:2005ie} in the context of the phase diagram of quenched Wilson ChPT. 
We find that both criteria for the choice of the appropriate saddle points are fulfilled by $\phi_5=-\phi_6=\omega_0$.
In this way one obtains the expected result given in eq.~\eqref{eq:grstate}.

We now repeat this procedure including the NLO lattice artefacts. We consider a ground state $u_{V,\;NLO}$ of the form given in eq.~\eqref{eq:vac} and we perturb the LO solution by choosing
\begin{eqnarray}
\phi_1 & = & \phi_2=\delta_s, \\
\phi_3 & = & \omega_0+\delta_v, \;\;\phi_4=-(\omega_0+\delta_v),\\
\phi_5 & = &  \omega_0+\delta_g,\;\; \phi_6=-(\omega_0+\delta_g),
\end{eqnarray}
where $\delta_{s,v,g}$ are corrections of $O(p^2)$.
We then perform the same analytic continuation on the ghost components,
\begin{equation}
\phi_5=i\hat{\phi_5}=i(\omega_0+\hat{\delta}_g),\;\;\;\phi_6=i\hat{\phi_6}=-i(\omega_0+\hat{\delta}_g).
\end{equation} 
By expanding the NLO potential at $O(p^4)$ we obtain
\begin{eqnarray}
V_{NLO}=a_0+a_1\delta_s+ a_2\delta_s^2+a_3\delta_v+ a_4\delta_v^2+a_5\hat{\delta}_g+a_6\hat{\delta}_g^2+\ldots=V_{q,NLO}+V_{g,NLO} ,
\end{eqnarray} 
with
\begin{eqnarray}
a_0 & = &  -2 \left(2 \hat{a}^2 (4 {W_6'}+{W_8'})+{m} (4 \hat{a} B (4 {W_6}+{W_8})+{\Sigma})\right),\\
a_1 & = & 0 \nonumber\\
a_2 & = & m\Sigma,\nonumber\\
a_3 & = & ia_5= 16 \hat{a} \sin w_0  (\hat{a} {W_8'} \cos w_0 +2 \hat{a} {W_6'}+B {m_P} {W_8}+2 B {m} {W_6}),\nonumber\\
a_4 & = & a_6 =m_P\Sigma.\nonumber
\end{eqnarray}
Also in this case the potential can be split in a real (sea + valence) part $V_{q,NLO}$ and a complex (ghost) part $V_{g,NLO}$.
The saddle point expansion yields
\begin{equation}
\delta_s =  0,\;\;\; \delta_v =  -\frac{a_3}{2a_4},\;\;\;\hat{\delta}_g  = -\frac{a_5}{2a_6}= -i\delta_v.
\end{equation}
The ground state of the theory at NLO is then given by
\begin{equation}
u_{V,\; NLO}=(\underbrace{1}_{2\times 2\; {\rm sea}}, \underbrace{e^{i\tau^3(\omega_0+\delta)/2}}_{2\times 2\; {\rm val}},\underbrace{e^{i\tau^3(\omega_0+\delta)/2}}_{2\times 2\; {\rm ghost}}),
\end{equation}
with
\begin{equation}\label{eq:delta}
\delta= \delta_v=\delta_g=-\frac{16\hat{a}\sin w_0}{F^2}\left\{\left(  W_6\cos w_0 +\frac{W_8}{2}\right)+\frac{2\hat{a}}{M_{vv}^2}\left(W_6'+\frac{W_8'}{2}\cos w_0  \right)          \right\}.
\end{equation}
Since the discretisation effects represent a perturbation in our power counting, this solution does not impose constrictions on the absolute value or the sign of the LECs $W_6,W_8,W_6',W_8'$.

\subsection{Spectral density in infinite volume}
The spectral density can be computed from $\langle\mathcal{P}^3\rangle$ by taking the discontinuity along imaginary $\mu_v$, as given  in eq.~\eqref{eq:disc_latt}.
At LO we obtain the known result
\begin{equation}\label{eq:rhoQ_lo}
[\rho_Q(\lambda,m)+\rho_Q(-\lambda.m)]_{LO}=\frac{2\Sigma\lambda}{\pi\sqrt{\lambda^2-m^2}},
\end{equation}
which does not depend on the volume and on the lattice spacing in our specific regime.
The NLO result in infinite volume is given by
\begin{eqnarray}
 [\rho_Q(\lambda,m)+\rho_Q(-\lambda,m)]_{NLO} &=& \frac{2\Sigma\lambda}{\pi\sqrt{\lambda^2-m^2}}\Bigg\{1+\frac{m^2\tilde\Delta}{\lambda^2-m^2}+  \frac{\Sigma}{(4\pi)^2F^4}\Bigg[-\pi\sqrt{\lambda^2-m^2}\nonumber\\
& +& m(3\bar{L}_6-1)+2\sqrt{\lambda^2-m^2}\arctan \left(\frac{\sqrt{\lambda^2-m^2}}{m} \right)\label{eq:rhoQ_nlo}\\
& - & 2m\ln\left(\frac{\Sigma|\lambda|}{F^2\mu^2} \right) -m\ln\left(\frac{2\Sigma\sqrt{\lambda^2-m^2}}{F^2\mu^2}   \right)\Bigg]+\frac{16\hat{a}}{F^2}W_6\Bigg\},\nonumber
\end{eqnarray}
where $\tilde\Delta$ arises from the correction to the ground state and is given by
\begin{equation}\label{eq:Delta_rho}
\tilde\Delta=\frac{16\hat{a}}{F^2}\left(W_6+\frac{2\hat{a}W_6'}{M^2_{ss}}   \right).
\end{equation}
If we define a shifted sea quark mass as follows
\begin{equation}\label{eq:masshat}
\hat{m}=m\left(1+\tilde\Delta\right),
\end{equation}
under the condition that $\lambda\gg \hat{m}$, the correction proportional to $\tilde{\Delta}$ can be resummed.
In this case one could then rewrite eq.~\eqref{eq:rhoQ_nlo} by omitting the term proportional to $\tilde\Delta$ and substituting $m\rightarrow \hat{m}$ everywhere. This resummed formula would be equivalent to the one in eq.~\eqref{eq:rhoQ_nlo} up to higher order corrections.\\
Notice that the term $\hat{a}/M^2_{ss}$ in eq.~\eqref{eq:Delta_rho} is not singular in the GSM regime.
The UV divergences arising from $G^1(0,M^2)$ and $G^2(0,M^2)$ have been cancelled by defining renormalised couplings
\cite{Gasser:1983yg}. Using dimensional regularisation in $4-2\epsilon$ dimensions and adopting the convention of \cite{Giusti:2008vb} we define
\begin{equation}
L_6=\frac{3\mu^{-2\epsilon}}{64(4\pi)^2}\left\{\bar{L}_6-\frac{1}{\epsilon}+\gamma-\ln 4\pi-1   \right\},
\label{eq:l6}
\end{equation}
where $\mu$ is a renormalisation scale. 

\begin{figure}
\includegraphics[width=12cm]{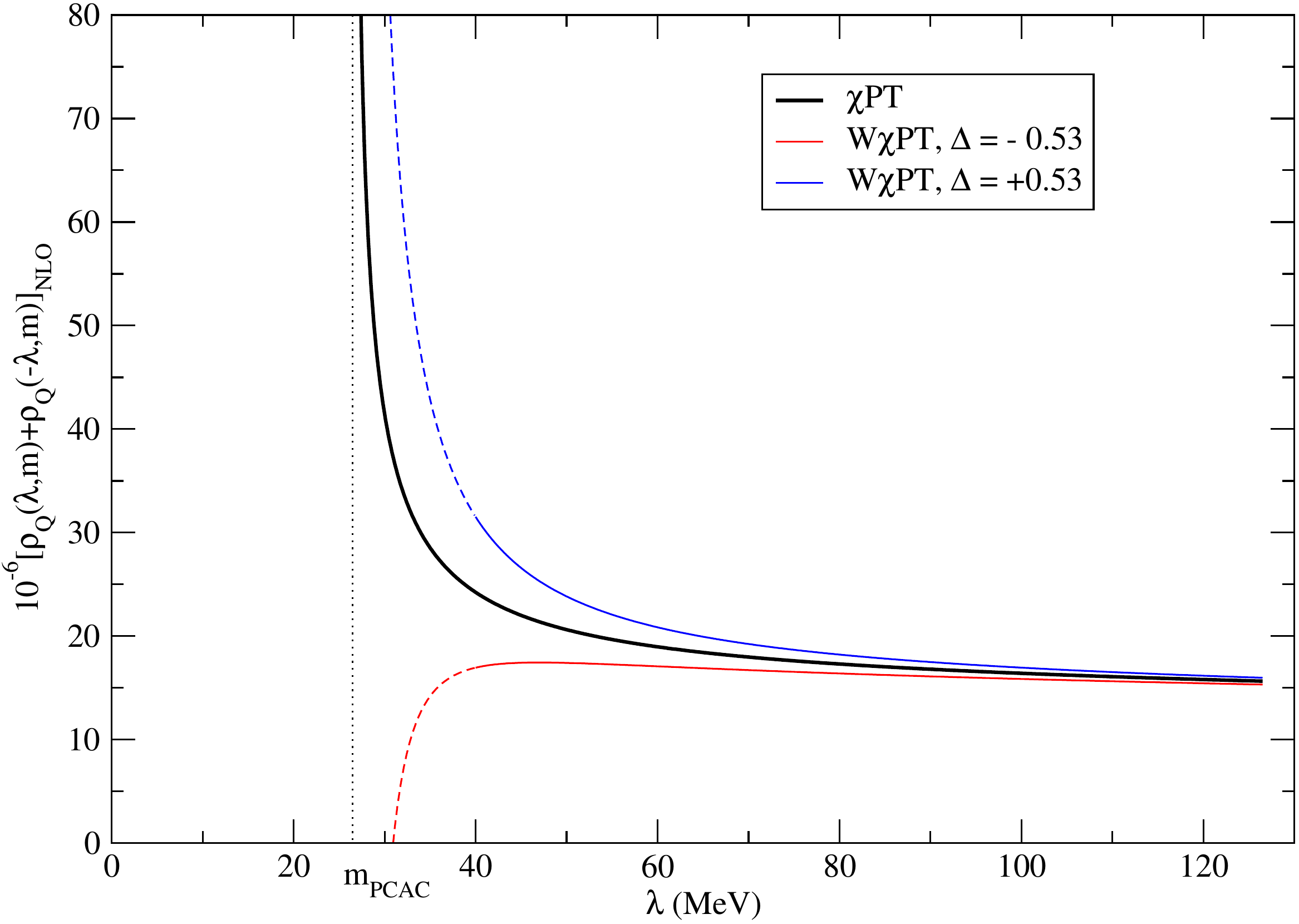}
\caption{The spectral density $[\rho_Q(\lambda,m)+\rho_Q(-\lambda,m)]_{NLO}$ in the infinite volume. 
We used the parameters $\Sigma=(275\;{\rm MeV})^3$, $m_{\rm PCAC}=26.5$ MeV,$F=90$ MeV, $\bar{L}_6=5$, $\mu=139.6$ MeV. 
The solid black line corresponds to the continuum $\chi$PT prediction, while 
the red (blue) lines correspond to the lattice W$\chi$PT prediction ($O(a)$-improved) on eq.~\protect\eqref{eq:rhoQ_nlo} with $\hat{a}^2W'_8= \pm 5\cdot 10^6\; {\rm MeV}^4$, 
corresponding to $\Delta=\mp 0.53$.} \label{fig:rhoNLO}
\end{figure}

The integrated spectral density $N(\Lambda_1,\Lambda_2,m)$ defined in eq.~\eqref{eq:modenum} can then be computed straightforwardly. For better readability
we report it in App. \ref{sec:app_c}, eq.\eqref{eq:int_rho}.
We stress that in the continuum these expressions coincides with the ones computed in \cite{Giusti:2008vb}.\\
It is useful to express our results for the spectral density in terms of the PCAC quark mass.  The NLO lattice corrections to the PCAC mass can be computed in the conventional W$\chi$PT for $N_f=2$.  The result is \cite{Sharpe:2004ny}
\begin{equation}
m_{\rm PCAC}= m\left\{ 1+\frac{16\hat{a}}{F^2} \left(W_6+\frac{W_8}{2} +\frac{W_{10}}{4} +\frac{2\hat{a}}{M^2_{ss}}\left(W_6'+\frac{W_8'}{2}  \right)      \right) \right\},
\label{eq:mpcac}
\end{equation}
where $W_{10}$ is an extra LEC. Eq.~\eqref{eq:rhoQ_nlo} can be then rewritten by substituting everywhere $m\rightarrow m_{\rm PCAC}$ and $\tilde\Delta\rightarrow\Delta$, with 
\begin{equation}
\Delta=-\frac{16\hat{a}}{F^2} \left(\frac{W_8}{2} +\frac{W_{10}}{4} +\frac{\hat{a}W_8'}{M^2_{ss}}  \right).
\label{eq:Delta}
\end{equation}
As before, the correction given by $\Delta$ can be resummed; the shift in the quark mass given in eq.~\eqref{eq:masshat} can be rewritten as a shift in the PCAC mass:  
\begin{equation}\label{eq:m_mpcac}
\hat{m}=m\left(1+{\tilde\Delta}\right)=   m_{\rm PCAC}\left(1+\Delta\right).    
\end{equation}
Notice that at NLO we can consistently substitute $m\rightarrow m_{\rm PCAC}$ in $M^2_{ss}$.\\
In fig.~\ref{fig:rhoNLO} we plot $[\rho_Q(\lambda,m)+\rho_Q(-\lambda,m)]_{NLO}$ in the continuum theory and in the discretised case with $W_6=W_8=W_{10}=0$, corresponding to a non-perturbatively $O(a)$-improved theory. We consider in particular a quark mass $m_{\rm PCAC}=26.5$ MeV;
the values of other parameters used in this plot are specified in the caption and justified by our numerical analysis
discussed in sect.~\ref{sec:disc}.\\
The central black curve corresponds to the continuum case; $W'_8>0$ ($\Delta<0$) gives rise to 
the red curve, while $W'_8<0$ ($\Delta>0$)  corresponds to the blue curve. 
The dashed lines for $\lambda\lesssim 40$ MeV are a reminder that our result 
is not expected to be valid close to the threshold.  
This issue will be discussed in more details in sec. ~\ref{sec:disc}.

\subsection{Finite volume corrections}\label{sec:fse}
Finite-volume effects arise at NLO and can be computed by taking into account the corrections to the propagators given in eq.~\eqref{eq:fsprop}. 
We can write down the spectral density as sum of the infinite volume and finite-volume corrections:
\begin{equation}
\rho_Q^V(\lambda,m)+\rho_Q^V(-\lambda,m)=\rho_Q(\lambda,m)+\rho_Q(-\lambda,m)+\Delta\rho_Q^V(\lambda,m)+\Delta\rho_Q^V(-\lambda,m).
\label{eq:rho_FSE1}
\end{equation}
The explicit expression for the finite-volume correction $\Delta\rho_Q^V(\lambda,m)+\Delta\rho_Q^V(-\lambda,m)$ is given in App.~\ref{sec:app_c}, eq.~\eqref{eq:rho_FSE}.
By integrating over $\lambda$ we obtain the corresponding finite-volume corrections for the integrated density
\begin{equation}
N^V(\Lambda_1,\Lambda_2,m)=N(\Lambda_1,\Lambda_2,m)+\Delta N^V(\Lambda_1,\Lambda_2,m).
\label{eq:N_FSE1}
\end{equation}
In eq.~\eqref{eq:N_FSE} we give the formula for $\Delta N^V(\Lambda_1,\Lambda_2,m)$ for the particular case $\Lambda_1=m$.

\begin{figure}
\includegraphics[width=12cm]{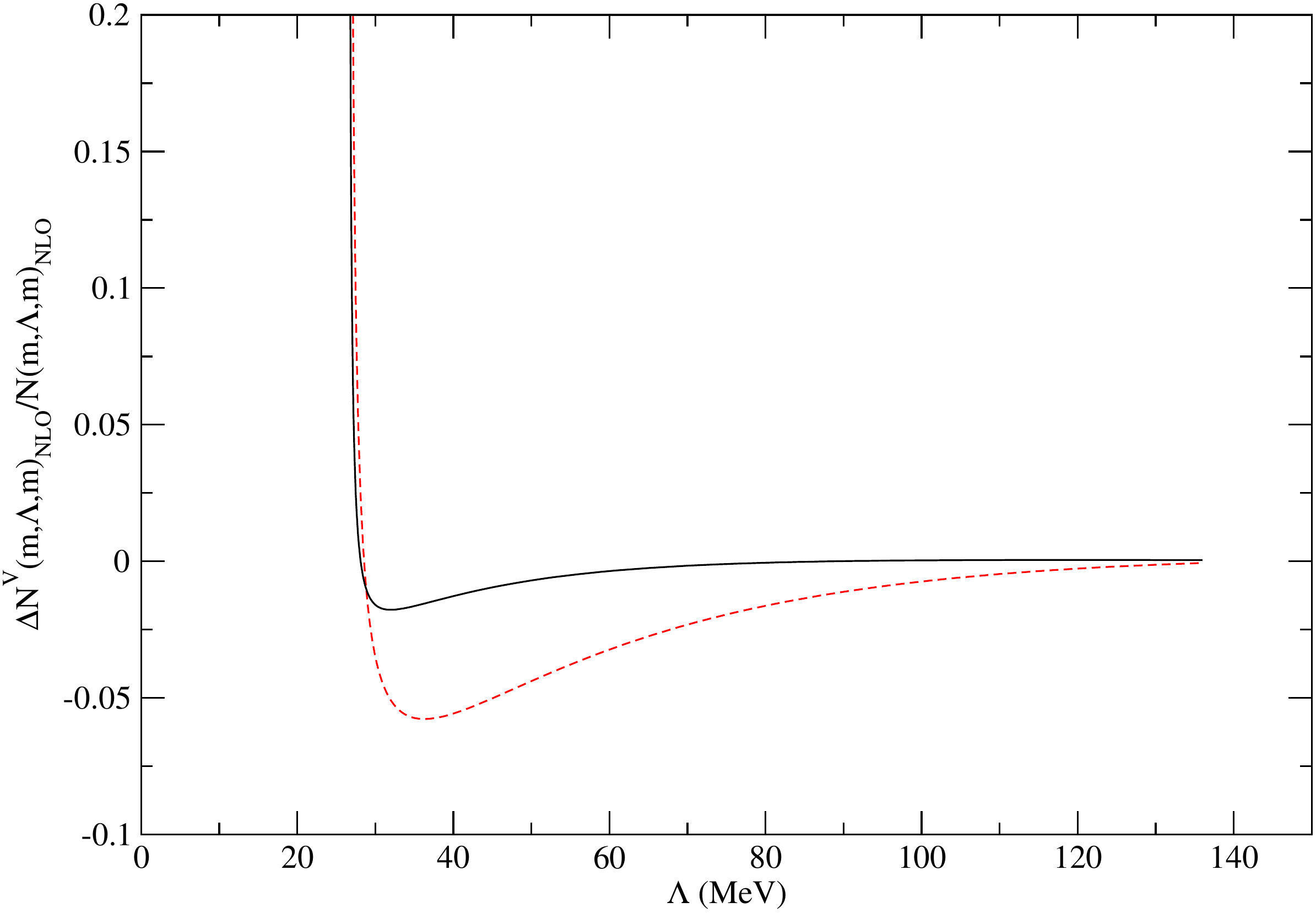}
\caption{The finite-size effects for the (continuum) integrated spectral density, $\Delta N^V(m,\Lambda,m)_{NLO}/N(m,\Lambda,m)_{NLO}$ as a function of $\Lambda$. We used the parameters $\Sigma=(275\;{\rm MeV})^3$, $m=m_{\rm PCAC}=26.5$ MeV, $F=90$ MeV, $\bar{L}_6=5$, $\mu=139.6$ MeV. The black solid line corresponds to $L=2.56$ fm, $T=3.84$ fm, while the red dashed line corresponds to  $L=1.92$ fm, $T=3.84$ fm.} \label{fig:modenum_fv}
\end{figure}

We stress that in our power counting finite-volume effects are insensitive to lattice artefacts at NLO, and these results are the one of the continuum theory already obtained in \cite{Giusti:2008vb}.\footnote{The formulae for finite-volume effects are not given explicitly in \protect\cite{Giusti:2008vb}; we checked 
 nevertheless that our results are in agreement with them \protect\cite{Giusti:pv} and we decided to report them in the present work for completeness.}
In fig. \ref{fig:modenum_fv} we show the relative correction $\Delta N^V(m,\Lambda,m)_{NLO}/N(m,\Lambda,m)_{NLO}$ as a function of $\Lambda$, for  two different volumes with $T=3.84$ fm and $L=1.92$ (red dashed curve) and $L=2.56$ fm (black solid curve).\\
Notice that the finite-volume correction $N^V(\Lambda_1,\Lambda_2,m)$ diverges at the threshold ($\Lambda_1=\Lambda_2=m$): this could be interpreted as a signal that, even if the quark mass is in the $p$-regime, another scale comes into play in this problem, namely $\sqrt{\lambda^2-m^2}$, and it needs to be treated with a different power counting when approaching the threshold at finite volume. 
Since in the $\epsilon$-expansion of the chiral theory divergences naturally appear when taking the chiral limit at finite volume, this might be the correct power counting to adopt for this additional scale when  $\sqrt{\lambda^2-m^2}\Sigma V\sim 1$. Having said that, for $L\gtrsim 2.0$ fm and sufficiently far from the threshold, finite-volume corrections amount to few percents.

\section{Discussion of the results and conclusions}
\label{sec:disc}
The important results of this paper are the expression of the spectral
density eq.~\eqref{eq:rhoQ_nlo} and its integrated form~\eqref{eq:int_rho}.
They are both obtained in the $p$-regime of W$\chi$PT at NLO.
Eq.~\eqref{eq:rhoQ_nlo} shows that the spectral density, when computed on a lattice with Wilson fermions,
is modified in two aspects with respect to the NLO continuum formula 
(see for example refs.~\cite{Damgaard:1998xy,Giusti:2008vb} or simply set all the $W$s 
in eq.~\eqref{eq:rhoQ_nlo} to zero).
The lattice artefacts modify the behaviour of the spectral density near the threshold, i.e. the term
proportional to $\tilde{\Delta}$ in~\eqref{eq:rhoQ_nlo}, and 
and its absolute normalisation, i.e. the term proportional to $W_6$ in~\eqref{eq:rhoQ_nlo}. 
The absolute normalisation correction proportional to $W_6$ vanishes if the theory is 
non-perturbatively O($a$)-improved. If the theory is not improved, beside the term proportional to $W_6$,
the correction to the normalisation contains in principle an additional O($a$) term stemming 
from the renormalisation constant $Z_P$.
We prefer to omit in our final formula~\eqref{eq:rhoQ_nlo} 
this term because it strictly depends on the way $Z_P$ is computed and only in very 
few cases we can have a suitable representation in W$\chi$PT~\cite{Aoki:2009ri}. 
We stress nevertheless that when analysing the spectral density computed in the unimproved theory
using eq.~\eqref{eq:rhoQ_nlo} the O($a$) cutoff effects of $Z_P$ have to be taken into account in some way.

More interestingly the cutoff effects modify, the $\lambda$ and $m$ dependence of the continuum formula.
These corrections, expressed in terms of the PCAC mass (see eqs.~(\ref{eq:mpcac}-\ref{eq:m_mpcac})), depends
on the following LECs: $W_8$, $W_{10}$ and $W_8'$. 
The modifications induced by cutoff effects to the continuum formula are plotted in fig.~\ref{fig:rhoNLO}.
To simplify the discussion we have chosen a lattice setup where the theory has been 
non-perturbatively improved, thus we set all the O($a$) LECs $W_{6}=W_{8}=W_{10}=0$. 
With the black curve we plot the $\lambda$ dependence of the spectral density 
in the continuum as given by NLO $\chi$PT~\cite{Damgaard:1998xy,Giusti:2008vb}
at given values for $m$, $\Sigma$ and $F$ (see the caption of the plot).
In the same plot we also show the modifications to the continuum formula induced by W$\chi$PT at NLO.
The first observation we make is that the way the spectral density depends on $\lambda$, close
to the threshold, is rather different from the continuum formula.
The second observation we make is that the W$\chi$PT formula has a rather different 
behaviour depending on the sign of $W_8'$, especially close to the threshold.
If $\Delta>0$ when $\lambda \rightarrow m_{\rm PCAC}$ the W$\chi$PT formula has a non-integrable singularity, 
while the continuum formula has an integrable singularity.
If $\Delta<0$ the spectral density goes to zero for a value of $\lambda >m_{\rm PCAC}$. These behaviours are shown
in fig.~\ref{fig:rhoNLO}.
This $\lambda$ dependence of the spectral density near the threshold is totally dominated
by the lattice NLO corrections induced by the vacuum realignment, indicating that our expansion cannot be trusted 
for too small $\lambda$. 
Our power counting breaks down near the threshold: more specifically, it is valid only when 
$\sqrt{\lambda^2-m^2}\sim a\Lambda_{\rm QCD}^2$, i.e. when the scale $\sqrt{\lambda^2-m^2}$ obeys a 
GSM power counting. By moving towards the threshold at finite lattice spacing one enters in the region where
 $\sqrt{\lambda^2-m^2}\sim a^2\Lambda_{\rm QCD}^3$ (Aoki regime), and lattice spacing corrections can not 
be treated in a perturbative way as it is done in this work.
In fig.~\ref{fig:rhoNLO} we draw dashed line where we assume our formula for the spectral
density to break down. The minimal $\lambda \simeq 40$ MeV is just an estimate obtained
fitting the available numerical data, as we will discuss futher on. It seems a reasonable estimate
given the fact that for $\lambda \simeq 40$ MeV the relative lattice spacing NLO corrections are still 
sufficiently small. Of course this is just an estimate based on the value of
$\Delta$ we input, i.e. from the size of the lattice artefacts. 
Only a thorough analysis of the numerical data can determine the value of $\Delta$, thus the range of validity
of our results.

Thus we can try to use our formula~\eqref{eq:int_rho} to fit the available numerical data 
and see if we obtain reasonable estimates of $\Sigma$ and $\Delta$.
To compare our formula with the available numerical data from ref.~\cite{Giusti:2008vb}
we decided to integrate the spectral density from $\lambda=\Lambda_1$ and $\lambda=\Lambda_2$,
taking $\Lambda_1 = 40$ MeV, in order to avoid the small $\lambda$ region where the lattice
corrections to the $\chi$PT formula dominate.
In order to compare with the results of~\cite{Giusti:2008vb} we use the mode number defined as 
$\nu(\Lambda_1,\Lambda_2,m)=V N(\Lambda_1,\Lambda_2,m)$ and we subtract the value $\nu(0,\Lambda_1,m)$
from the numerical data.
%
%
\begin{figure}[t]
 \includegraphics[width=12cm]{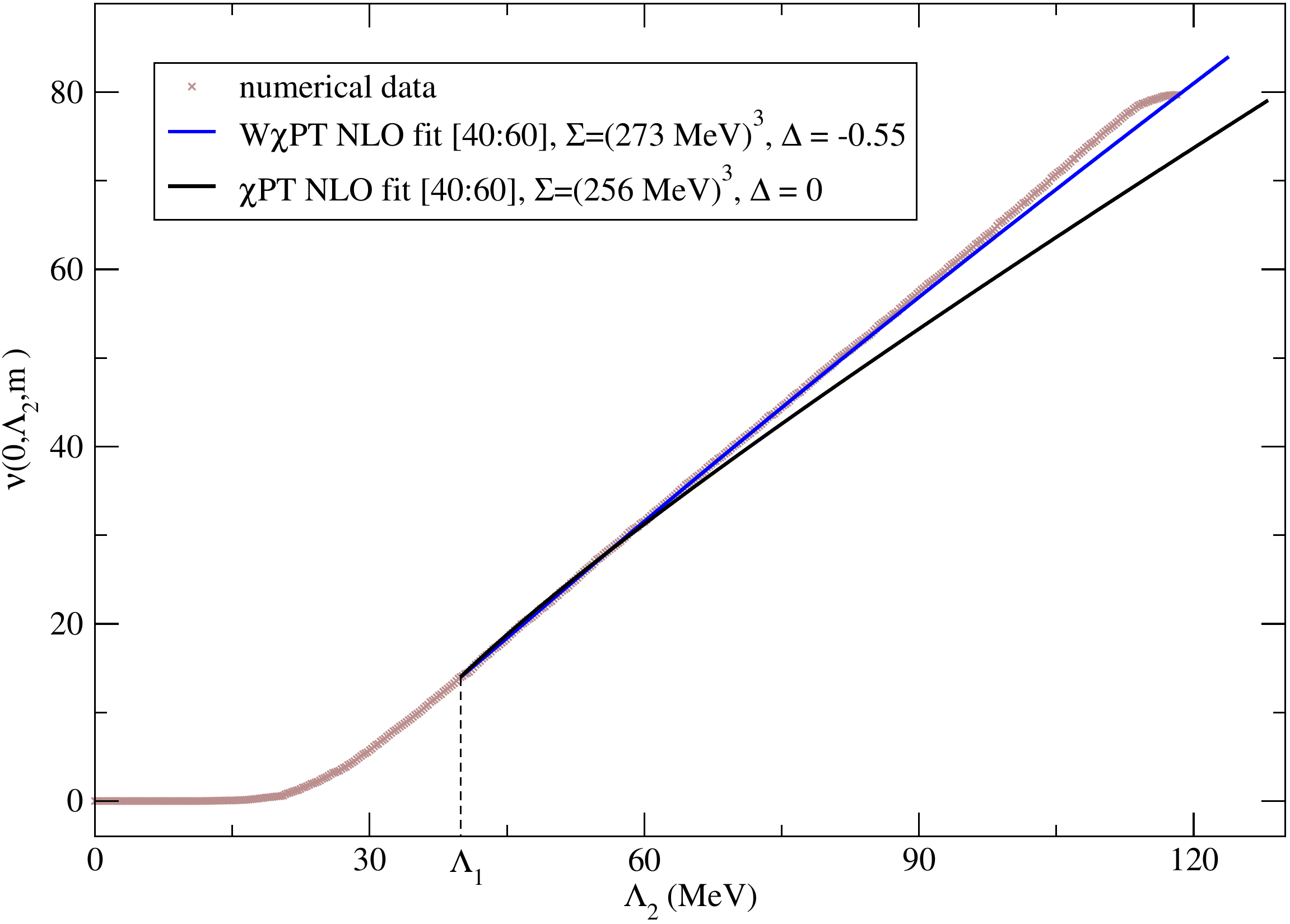}
\caption{Comparison of the numerical data (red curve) for the mode number $\nu(0,\Lambda_2,m)$ 
and the NLO fit results using continuum $\chi$PT (black curve) and W$\chi$PT (blue curve). 
The fits are obtained using the following parameters: $\Lambda_1 = 40$ MeV
$\mu=139.6$ MeV, $F = 90$ MeV, $\bar{L}_6 = 5$ and $m_{\rm PCAC}=26.5$ MeV. 
The fit range is $\Lambda_2 \in \left[40:60\right]$ MeV.
}\label{fig:mode_nlo}
\end{figure}
%
%
\begin{figure}
 \includegraphics[width=12cm]{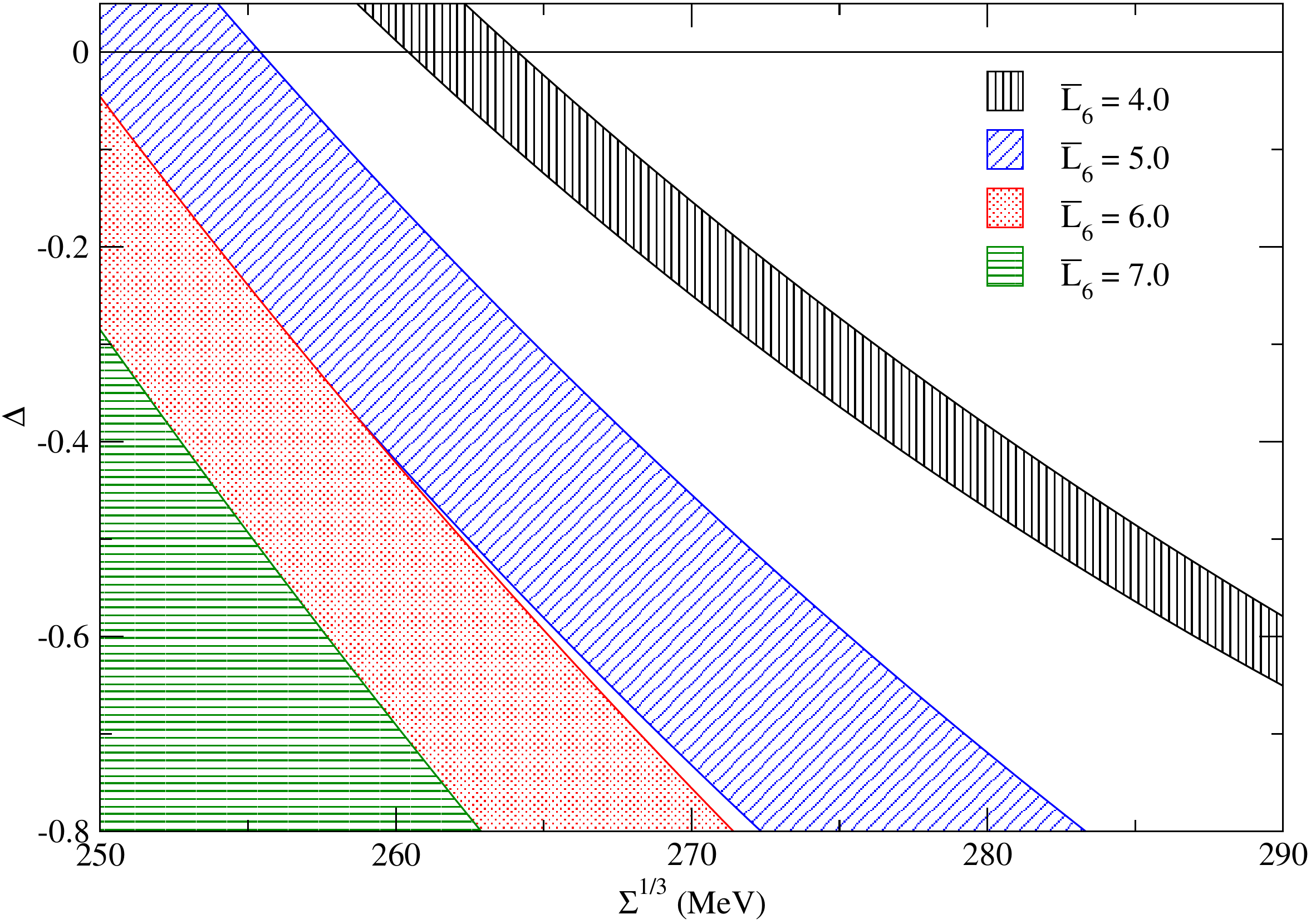}
\caption{Parameter space of the results obtained fitting the numerical data of 
ref.~\cite{Giusti:2008vb} using eq.~\eqref{eq:int_rho}. See the main text for details on the fit performed.
Only results where a reasonable fit is obtained are showed. }
\label{fig:Delta_Sigma}
\end{figure}
In fig.~\ref{fig:mode_nlo} we show a typical fit of the numerical data of 
ref.~\cite{Giusti:2008vb} using the infinite volume formul{\ae} of continuum $\chi$PT and W$\chi$PT in the range
$\Lambda_2 \in \left[40:60 \right]$ MeV.
The numerical data are obtained at a value of the renormalised PCAC mass $m_{\rm R}\simeq 26$ MeV. \\
We have performed fits using the both the NLO formul{\ae} of $\chi$PT and W$\chi$PT. \footnote{A LO fit with 
$\chi$PT does not perform better in any range of $\Lambda$ where the numerical data are available.}
With the continuum $\chi$PT formula we have fit $\Sigma$ and fixed the following parameters:
$\mu=139.6$ MeV, $F = 80,90$ MeV, $\bar{L}_6 = 3,4,\ldots ,7$ and $m_{\rm PCAC}=26.5$ MeV.
With the W$\chi$PT formula we have set all the O($a$) LECs $W_{6}=W_{8}=W_{10}=0$ 
because the numerical data are non-perturbatively O($a$)-improved. Additionally we have fixed the
value of $\Sigma^{1/3}$ within the range $250-300$ MeV and we have fit $\Delta$.
We have performed fits with several forms for the W$\chi$PT formula, all differing by NNLO
terms. We have observed that the best fits are obtained with eq.~\eqref{eq:int_rho}.
In order to decide on the 'best fits' we have selected the fit results that gives small values of the 
squared differences between the input data points and the theoretical formula evaluated at the same points.

A summary of the fit results, satisfying our criterion is given in fig.~\ref{fig:Delta_Sigma}. 
The plot shows the parameter space of the fit results in the $\Delta$, $\Sigma^{1/3}$ plane. Each coloured 
band represents a fixed value of $\bar{L}_6$ and the lower and upper borders of each band represent
respectively $F=80$ MeV and $F=90$ MeV. 
The two most important observations are that reasonable fits are obtained only for $4 \lesssim \bar{L}_6 \lesssim 7$ 
and for almost the whole parameter space we obtain a negative value of $\Delta$. Our best fit result shown in 
fig.~\ref{fig:mode_nlo} gives $\Delta = -0.55$ with a value of 
$\Sigma^{1/3} = 273$ MeV, which is in the right ballpark. 
We will come back to the consequences of this result at the end of this section. \\We remark that while the fit results obtained using the continuum NLO formula are not 
disastrous always a better agreement is found using our W$\chi$PT result.\\
The W$\chi$PT formula is able to describe well the numerical data also in a larger $\Lambda_2$ range, i.e.
$\left[ 40:100\right]$ MeV, with reasonable values of the chiral condensate.\\
Even if our analysis is rather qualitative and more numerical data are needed,
we can try to draw some conclusions.
W$\chi$PT describes the numerical data better than continuum $\chi$PT in the range of applicability
of our results, i.e. sufficiently away from the threshold, but we cannot exclude that adding NNLO terms
in the continuum formula would improve the quality of the fits.

Our fit results seem to prefer a negative value for $\Delta$ implying a positive value for
$W_8'$. 
In the literature, the LECs $W'_6$ and $W'_8$ are often found combined as \cite{Aoki:2004ta}
\begin{equation}
\hat{a}^2\left(W'_6+\frac{W'_8}{2}  \right)\equiv -\frac{F^2c_2a^2}{16},
\label{eq:c2}
\end{equation}
which is the combination appearing in quantities computed in full $N_f=2$ W$\chi$PT. 
The chiral phase diagram for Wilson-type fermions depends on the LEC $c_2$~\eqref{eq:c2} and in particular 
two scenarios can take place depending on its sign.
In our case only one of the two LECs appears in the discretisation correction to the spectral density.  
For large number of colours $N_c$, one finds $W_6'/W_8'\sim 1/N_c$ \cite{Sharpe:2006ia}, 
indicating that $W_8'$ is the dominant contribution in the coefficient $c_2$. 
This implies that independently on the sign of $W_6'$ a positive value for $W_8'$ would 
suggest a negative value for $c_2$. \footnote{While for other discretisations existing numerical results of 
refs.~\cite{Farchioni:2004us,Farchioni:2004fs,Farchioni:2005tu} are consistent with $c_2<0$, 
for non-perturbative O($a$)-improved Wilson fermions and Wilson gauge action the value of $c_2$ is still unknown.}
Assuming for instance $W'_6=W'_8/3$, then the results we obtain from the fit $\Delta=-0.55$
imply that $\hat{a}^2 W'_8=5.4\cdot 10^6\; {\rm MeV}^4$ which for $a=0.08$ fm 
would correspond to $c_2\simeq - (473 \;{\rm MeV})^4$. 
It would be interesting to extend the numerical data in order to establish more accurately
the sign of and the absolute value of $W_8'$. This will give us valuable informations
on the behaviour of the spectrum of the Hermitean Wilson operator and a hint on the
scenario for the chiral phase diagram that takes place with clover fermions and Wilson gauge action. \footnote{If the negative sign of $c_2$ found in this work will be confirmed by other studies, it will imply that the pion mass never vanishes. With our numerical estimate of $c_2$, the minimal pion mass at $a=0.08$ fm would be around 130 MeV, which is well below the values actually simulated with this action.}

We close this section by stressing again an important remark. Our results seem to indicate that apart from the quark mass, 
the lattice spacing and the lattice size, 
there is an additional scale entering in this problem, namely the eigenvalues of the massless Dirac operator. 
In particular, after the analytic continuation of eq.~\eqref{eq:disc_latt} the mass $\sqrt{m^2+\mu_v^2}$ 
becomes $\sqrt{\lambda^2-m^2}$, which parametrizes the ``distance'' from the threshold of the spectral density.
Even if we assume, as it is done in this work, that all masses are in the $p-$regime the value of this
additional scale can be arbitrarily small. In order to get meaningful results we have to 
implicitly assume that this parameter obeys the same power counting as the quark masses with 
respect to the lattice spacing $a$ and the lattice size $L$. 
While this fact seems trivial at this stage, it is not totally obvious to infer it from the partially quenched initial setup, 
where valence quarks are introduced as probes to obtain the spectral density. This implies that, at fixed $a$ and $L$, 
our results will not be valid in the vicinity of the threshold, since from one side the $p$-regime 
expansion will fail, and from the other side the GSM counting will not be valid anymore, because close to the threshold,
as we have discussed before, the NLO corrections induced by the finite lattice spacing dominate the LO result.

We stress that these two effects are decoupled: close to the threshold, there is a scale which in 
continuum $\chi$PT at finite volume needs to be treated with the $\epsilon$-expansion, 
and which in W$\chi$PT in infinite volume needs to be treated with the Aoki power counting.
The way the two effects combine close to threshold is a not trivial issue and it deserves further 
investigations.

We conclude that our formula is a useful tool, combined with numerical data,
to understand the behaviour of the spectral density of the Hermitean Wilson 
operator. We also consider this work a necessary and important step towards a complete 
theoretical understanding of the behaviour close to the threshold of the spectral density 
of the Hermitean Wilson operator.

\section{Acknowledgments}
We are indebted to Leonardo Giusti for the access to some crucial private notes on the graded-group method
for partially quenched Chiral Effective Theory and for useful discussions. We thank Steve Sharpe 
and Leonardo Giusti for a
critical reading of the manuscript and for precious comments,
as well as Martin L\"uscher for advices and suggestions and for providing numerical 
data for the spectral density of the Wilson Dirac operator.

\newpage

\begin{appendix}
\section{SU($m|n$) graded group: conventions and properties}
\label{sec:app_a}
In this Appendix we collect conventions and properties related to SU($m|n$) \cite{Cornwell:1989bx,Giusti:pv} which are relevant for the partially quenched Chiral Effective Theory in the graded-group formulation. 
\subsection{Supermatrices}
For $n,m$ positive integers, a square even supermatrix is defined as a $(m+n)\times (m+n)$ matrix with the structure
\begin{equation}\label{eq:supermat}
U=\left(
\begin{array}{ll} A_{m\times m} & B_{n\times m} \\
C_{m\times n} & D_{n\times n}.
\end{array}
\right),
\end{equation}
where $A$ and $D$ have elements in the \emph{even} subspace of the Grassmann algebra, while $B$ and $C$ have elements belonging to the \emph{odd} subspace of the Grassmann algebra (see \cite{Cornwell:1989bx} for definitions and properties of the Grassmann algebras).
We define the \emph{supertrace} of a supermatrix
\begin{equation}
\str{U}\equiv \tr{A}-\tr{D},
\end{equation}
and the \emph{superdeterminant} of an invertible supermatrix as
\begin{equation}
\sdet{U} \equiv  \Det{A-BD^{-1}C}/\Det{D}.
\end{equation}

\subsection{Superalbegras}
We now consider a square  $(m+n)\times (m+n)$ matrix with complex entries and the structure
\begin{equation}
M=\left(
\begin{array}{ll} A_{m\times m} & B_{n\times m} \\
C_{m\times n} & D_{n\times n}.
\end{array}
\right).
\end{equation}
$M$ is said to be \emph{even} if $B=C=0$, whereas it is \emph{odd} if $A=D=0$. The degree or parity of $M$ is defined to be 
\begin{equation}
{\rm deg}\;M =\left\{
\begin{array}{ll}
0 & {\rm if}\; M\;  {\rm is\; even}\\
1 & {\rm if}\; M\;  {\rm is\; odd}
\end{array}
\right.
\end{equation}
One can show that set of all complex linear combinations of these matrices form an associative superalgebra. The \emph{supertrace} can be defined analogously as in the case of supermatrices,
\begin{equation}
\str{M}\equiv \tr{A}-\tr{D}.
\end{equation}

\subsection{Lie supergroups and superalgebras}
The Lie Supergroup SU($m|n$) is the set of $(m+n)\times (m+n)$ even supermatrices $U$
satisfying the conditions 
\begin{equation}
U^\dagger U =  1,\;\;\;\;\sdet{U} =  1.
\end{equation}
By parametrizing $U$ as
\begin{equation}
U\equiv e^{i\phi},
\end{equation}
these conditions convert into
\begin{equation}
\phi = \phi^\dagger,\;\;\;\;\str{\phi}  =  0.
\end{equation}
The Lie superalgebra associated to SU($m|n$) can be constructed by supplementing the associative superalgebra of $(m+n)\times (m+n)$ complex matrices defined in the previous subsection with the generalised commutators 
\begin{equation}
[M,N]=MN-(-1)^{({\rm deg}\;M)({\rm deg}\;N) } NM,
\end{equation} 
and by requiring the conditions
\begin{equation}
M = M^\dagger,\;\;\;\;\str{M}  =  0.
\end{equation}
A basis for the Lie superalgebra of SU($m|n$) can be chosen to be a set of Hermitian matrices $T^a=(T^a)^\dagger$ $(a=1,\ldots,(m+n)^2-1)$, with a definite parity
\begin{equation}
{\rm deg}(a)\equiv {\rm deg}T^a,
\end{equation}
and zero supertrace
\begin{equation}
\str T^a=0.
\end{equation}
The normalisation is given by 
\begin{equation}
\str{T^aT^b}=\frac{g^{ab}}{2},
\end{equation}
where
\begin{equation}\label{eq:gab}
g^{ab}=\left(
\begin{array}{lllllllll}
1 &  & & & &     &&&\\
  & \ddots & & & &&&&\\
  & & 1 &&&&&&\\
& & & -\tau^2 & &&&& \\
& & & & \ddots & &&& \\
& & & & & -\tau^2  &&& \\
& & & & &  & -1 && \\
& & & & &  &  & \ddots &\\
& & & & &  & && -1 
\end{array}
\right)
\begin{array}{ll}
\Bigg\}& 1,\ldots,m^2-1\\
& \\
\Bigg\} & m^2,\ldots, m^2+2mn-1\\
& \\
\Bigg\} & m^2+2mn,\ldots, (m+n)^2-1,
\end{array}
\end{equation}
with $g^{ab}=(-1)^{{\rm deg}(a){\rm deg}(b) }g^{ba}$. The elements from $m^2$ to $m^2+2mn-1$ are odd matrices, while the remaining ones are even. 

In this work we consider the particular case of SU(4$|$2). 
With our conventions, $T^1,\ldots,T^{15}$ are the (even) generators of the SU(4) subgroup that includes sea and valence quarks, $T^{32},\ldots,T^{34}$ (even) are associated with the SU(2) ghost sector, $T^{16},\ldots,T^{31}$ (odd) mix the ghosts and the quarks, and finally $T^{35}$ (even) is a diagonal matrix with components in both quark and ghost sectors. 

For the computation of the propagator it is convenient to introduce the tensors
\begin{eqnarray}
k^{ab} & = & \left(\frac{1}{\sqrt{3}}g^{a14} +\frac{1}{\sqrt{6}}g^{a15}-\frac{1}{\sqrt{2}}g^{a35}  \right) \left(\frac{1}{\sqrt{3}}g^{b14} +\frac{1}{\sqrt{6}}g^{b15}-\frac{1}{\sqrt{2}}g^{b35}  \right)       ,\label{eq:kab}\\
h^{ab} & = & \left(\frac{1}{\sqrt{3}}g^{a14} +\frac{1}{\sqrt{6}}g^{a15}+\frac{1}{\sqrt{2}}g^{a35}  \right)\left(\frac{1}{\sqrt{3}}g^{b14} +\frac{1}{\sqrt{6}}g^{b15}+\frac{1}{\sqrt{2}}g^{b35}  \right)\label{eq:hab} 
\end{eqnarray}
which satisfy 
\begin{equation}
\sum_{c=1}^{35}h^{ac}k^{cb}=0.
\end{equation}

\section{The spectral density in W$\chi$PT with the replica method}
\label{sec:app_b}
In this appendix we present the same calculation done in sec. \ref{sec:density} applying the replica method \cite{Damgaard:2000gh} in alternative to the graded group method.
We consider a partially quenched theory with $N_s$ sea quarks of mass $m_s$ and $N_v$ quenched valence quarks of mass $m_v$. For simplicity we set the volume to infinity; finite-volume corrections can be computed in a standard procedure, like explained in sec. \ref{sec:fse}. 
The replica method consists in enlarging the valence sector to $k$ replica of $N_v$ valence quarks. The full symmetry group at zero quark mass is therefore SU$(N_s+N_r)_L$$\times$SU$(N_s+N_r)_R$, with $N_r=kN_v$, and the generating functional at the quark level is given by
\begin{equation}
Z_{PQ,replica}(J)=\int [dA_\mu]\Det{D+m_v+J}^{N_v}\Det{D+m_v}^{N_r-N_v}\Det{D+m_s}^{N_s}e^{-S_g(A_\mu)},
\end{equation}
 where we have introduced source terms $J$. In the limit $N_r=0$ one reproduces the generating functional in the graded group method  \cite{Bernard:1992mk,Bernard:1993sv}
\begin{equation}
Z_{PQ,graded}(J)=\int [dA_\mu]\frac{\Det{D+m_v+J}^{N_v}}{\Det{D+m_v}^{N_v}}\Det{D+m_s}^{N_s}e^{-S_g(A_\mu)},
\end{equation}
which is obtained in a SU$(N_s+N_v|N_v)_L$$\times$SU$(N_s+N_v|N_v)_R$ theory, where internal valence quark loops are cancelled by introducing quarks with the wrong statistics (ghosts).
This equivalence is translated into the chiral effective theory and has been verified at the perturbative level \cite{Damgaard:2000gh}.
In this appendix we explicitly show how this equivalence is realised for the computation of the spectral density starting from the partially quenched pseudoscalar density.

For our specific computation we consider the case $N_s=2$, with a doublet of twisted valence quarks. The chiral Lagrangian is the same as in eq.~\eqref{eq:fullL}, with the ``Supertrace'' replaced by the conventional trace over the group SU$(N_s+N_r)$. The mass matrix has the form 
\begin{equation}
\mathcal{M}={\rm diag} (\underbrace{m}_{2\times 2\; {\rm sea}}, \underbrace{\underbrace{m+i\mu_v\tau^3}_{N_v=2},\ldots,m+i\mu_v\tau^3}_{N_r=kN_v}).
\end{equation}
The pseudo Nambu-Goldstone fields are parametrised by
\begin{equation}
U(x)=u_Ve^{2i\xi(x)/F}u_V,\;\;\;\;\xi=\sum_a \xi^a T^a,
\end{equation}
where now $T^a,a=1,\ldots,(2+N_r)^2-1$ are the generators of SU$(2+N_r)$, with the normalisation convention
\begin{equation}
\tr{T^aT^b}=\frac{\delta^{ab}}{2},
\end{equation}
while $u_V$ represents the ground state of the theory. By minimising the LO potential we obtain
\begin{equation}\label{eq:vac_replica}
u_V={\rm diag} (\underbrace{1}_{2\times 2\; {\rm sea}}, \underbrace{e^{i\tau^3\omega_0/2}}_{N_r}),
\end{equation} 
with the definitions already given in eq.~\eqref{eq:omega_mp}
\begin{equation}
\sin\omega_0=\frac{\mu_v}{m_P},\;\;\;\cos\omega_0=\frac{m}{m_P},\;\;\;\;m_P=\sqrt{m^2+\mu_v^2}.
\end{equation}
In this framework it is convenient to write down the pseudo Nambu-Goldstone propagator in explicit components
\begin{equation}
\langle \xi_{ca}(x)\xi_{db} (y) \rangle=\frac{1}{2} \left[\delta_{cb}\delta_{da} G^1(x-y,M_{ab}^2)-\delta_{ca}\delta_{db}E(x-y,M_{aa}^2,M_{cc}^2)   \right],
\end{equation}
with $a,b,c,d=1,\ldots,(2+N_r)$.
The infinite-volume scalar propagator is defined in eq.~\eqref{eq:ginf}, while
\begin{eqnarray}
E(x-y,M_{aa}^2,M_{cc}^2) & \equiv & \frac{1}{(2\pi)^4}\int d^4p\frac{e^{ipx}}{(p^2+M_{aa}^2)(p^2+M_{cc}^2)F(p)}, \label{def_E}\\
F(p) & \equiv & \frac{2}{p^2+M_{ss}^2}+  \frac{N_v}{p^2+M_{vv}^2},
\end{eqnarray}
with
\begin{equation}
M_{ab}^2 =  B(m_a+m_b)=\left\{ 
\begin{array}{ll}
M_{ss}^2=2mB & {\rm if}\;\; a,b=1,2\\
M_{vv}^2=2m_PB & {\rm if}\;\; a,b=3,\ldots,N_r\\
M_{sv}^2=(m+m_P)B & {\rm if}\;\; a(b)=1,2;b(a)=3,\ldots,N_r.
\end{array}
\right.
\end{equation}
The observable we have to consider in order to extract the spectral density is the expectation value of the pseudoscalar density defined in eq.~\eqref{eq:pqcond}. 
The pseudoscalar density can be obtained by taking the functional derivative of the action with respect to appropriate sources, which are introduced by the following procedure:
\begin{equation}
\mathcal{M}\rightarrow \mathcal{M}+\hat{\tau}^3p^3(x),\;\;\;\;\mathcal{M}^\dagger\rightarrow \mathcal{M}^\dagger-\hat{\tau}^3p^3(x),
\end{equation}
where $\hat{\tau}^3$ has non-zero elements only in one of the replica of the valence sector
\begin{equation}
\hat{\tau}^3={\rm diag} (\underbrace{0}_{2\times 2\; {\rm sea}},\underbrace{\underbrace{\tau^3}_{N_v=2},0}_{N_r}).
\end{equation}
At LO we obtain the (continuum) expectation value
\begin{equation}
\langle\mathcal{P}^3\rangle_{LO}= 2i\Sigma\sin\omega_0.
\end{equation}
This result is independent on $N_r$, and coincide with the one obtained with the graded group method, eq.~\eqref{eq:p3lo}.
At $O(p^2)$ we obtain
\begin{eqnarray}
\langle\mathcal{P}^3\rangle_{NLO} & = &  2i\Sigma\sin(\omega_0)\Bigg\{1+\delta\cot\omega_0  -\frac{1}{F^2}\Bigg[2G^1(0,M^2_{sv})+N_rG^1(0,M^2_{vv}) \nonumber\\
&-& E(0,M^2_{vv},M^2_{vv})-  16L_6(2M^2_{ss}+N_rM^2_{vv})-4M^2_{vv}(H_2+2L_8)\\
&-& 8\hat{a}\left(W_6\left(2+N_r\cos\omega_0\right)+W_8\cos\omega_0\right)\Bigg]\Bigg\}\nonumber
\end{eqnarray}
The shift $\delta$ is the $O(p^2)$ correction to the ground state angle $\omega_0$ that must be computed by minimising the NLO potential.  Unlike the graded group case discussed in sec. \ref{sec:vac}, with the replica method the minimisation is trivial, since the fields belong to the conventional SU$(N_s+N_r)$ group.
By taking only the linear term in $N_r$ we get
\begin{equation}
\delta=-\frac{16\hat{a}\sin \omega_0}{F^2}\left\{\left(  W_6\cos \omega_0 +\frac{W_8}{2}\right)+\frac{2\hat{a}}{M^2_{vv}}\left(W_6'+\frac{W_8'}{2}\cos \omega_0  \right) \right\},
\end{equation}
which coincides with the shift calculated with the graded groups in sec. \ref{sec:vac}, eq.~\eqref{eq:delta}.
The final result in the limit $N_r\rightarrow 0$ for the partially quenched pseudoscalar density at NLO is then
\begin{eqnarray}
\langle\mathcal{P}^3\rangle_{NLO} &= & 2i\Sigma\sin\omega_0\Bigg\{1+\delta\cot\omega_0 + \frac{1}{F^2}\Bigg[\frac{1}{2}G^1(0,M^2_{vv})-2G^1(0,M^2_{sv})+32L_6M^2_{ss}\\
&-&\frac{1}{2}(M^2_{vv}-M^2_{ss})G^2(0,M^2_{vv})+4M^2_{vv}(H_2+2L_8) 
+ 8\hat{a}\left(2W_6+W_8\cos\omega_0\right)\Bigg]\Bigg\} \nonumber, 
\end{eqnarray}
where we have used 
\begin{equation}
E(0,M^2_{vv},M^2_{vv})|_{N_r\rightarrow 0} =\frac{1}{2}G^1(0,M^2_{vv})-\frac{1}{2}(M^2_{vv}-M^2_{ss})G^2(0,M^2_{vv}).
\end{equation}
This result for the pseudoscalar density is fully equivalent to the one obtained with the graded groups, eq.~\eqref{eq:p3nlograd}, as expected.

\section{Some explicit formulae}
\label{sec:app_c}
In this Appendix we collect some explicit formulae which we omitted in the main text for clarity reasons.

\subsection{Integrated spectral density at NLO}
We report the integrated spectral density computed at NLO in W$\chi$PT. Starting from the result presented in eq.~\eqref{eq:rhoQ_nlo} for the spectral density, we obtain:
\begin{eqnarray}\label{eq:int_rho}
N(\Lambda_1,\Lambda_2,m)_{NLO} &\equiv & \int_{\Lambda_1}^{\Lambda_2}d\lambda\; [\rho_Q(\lambda,m)+\rho_Q(-\lambda,m)]_{NLO}=\\
& = & \frac{2\Sigma}{\pi}\Bigg\{\left(\sqrt{\Lambda_2^2-m^2}-\sqrt{\Lambda_1^2-m^2}\right)\left(1+\frac{16\hat{a}}{F^2}W_6\right) \nonumber\\
&-&\tilde\Delta m^2 \left(\frac{1}{\sqrt{\Lambda_2^2-m^2}}-\frac{1}{\sqrt{\Lambda_1^2-m^2}}  \right) \nonumber\\
& + & \frac{\Sigma}{(4\pi)^2F^4}\Bigg[-\frac{\pi}{2}(\Lambda_2^2-\Lambda_1^2)     
+  m(1+3\bar{L}_6)\left(\sqrt{\Lambda_2^2-m^2}-\sqrt{\Lambda_1^2-m^2}\right)\nonumber\\
& + & (\Lambda_2^2-2m^2) \arctan\left(\frac{\sqrt{\Lambda_2^2-m^2}}{m}\right)- (\Lambda_1^2-2m^2) \arctan\left(\frac{\sqrt{\Lambda_1^2-m^2}}{m}\right)     \nonumber\\
&-&m\sqrt{\Lambda_2^2-m^2}\left(2\log\left(\frac{\Sigma\Lambda_2}{F^2\mu^2} \right)+ \log\left(\frac{2\Sigma\sqrt{\Lambda_2^2-m^2}}{F^2\mu^2}  \right)  \right)\nonumber\\
&+&m\sqrt{\Lambda_1^2-m^2}\left(2\log\left(\frac{\Sigma\Lambda_1}{F^2\mu^2} \right)+ \log\left(\frac{2\Sigma\sqrt{\Lambda_1^2-m^2}}{F^2\mu^2}  \right)  \right)\Bigg]\Bigg\},\nonumber
\end{eqnarray}
where $\tilde\Delta$ is given in eq.~\eqref{eq:Delta_rho}.

\subsection{Finite volume corrections}
In this section we give explicit expressions for the finite volume corrections defined in eqs.~(\ref{eq:rho_FSE1}),(\ref{eq:N_FSE1}).
It is convenient to represent the finite-volume propagator $g_r(M^2)$ \cite{Hasenfratz:1989pk} defined in eq.~\eqref{eq:fsprop} as sum over modified Bessel functions $K_\nu$:
\begin{equation}\label{eq:fspropbes}
g_r(M^2)=\frac{1}{\Gamma(r)(4\pi)^2}\sum_{\{n1,n2,n3,n4\}\neq 0}F_{r-2}\left(\frac{q_n^2}{4},M^2 \right),
\end{equation}
with $q_n^2=\left((n_1^2+n_2^2+n_3^2)L^2+n_4^2T^2\right)$ and
\begin{equation}
F_\nu(a,z)=2\left(\frac{a}{z}  \right)^{\nu/2}K_\nu(2\sqrt{az}).
\end{equation} 
Using this representation one obtains the finite-volume correction to the spectral density 
\begin{eqnarray}
\label{eq:rho_FSE}
[\Delta\rho_Q^V(\lambda,m)+\Delta\rho_Q^V(-\lambda,m)]_{NLO} &= & \frac{\Sigma}{\pi}\frac{\Sigma}{F^4(4\pi)^2}\frac{2\lambda}{\sqrt{\lambda^2-m^2}} \sum_{\{n1,n2,n3,n4\}\neq 0}\\
& & 
\Bigg\{{\rm Re}\left[F_{-1}\left(\frac{\Sigma q_n^2}{2F^2},i\sqrt{\lambda^2-m^2}\right) \right]\nonumber\\
&-&
2{\rm Re}\left[F_{-1}\left(\frac{\Sigma q_n^2}{4F^2},m+i\sqrt{\lambda^2-m^2}\right) \right]\nonumber\\
&+&\sqrt{\lambda^2-m^2}{\rm Im}\left[F_{0}\left(\frac{\Sigma q_n^2}{2F^2},i\sqrt{\lambda^2-m^2}\right) \right]\nonumber\\
&+ & m{\rm Re}\left[F_{0}\left(\frac{\Sigma q_n^2}{2F^2},i\sqrt{\lambda^2-m^2}\right) \right]\Bigg\}.\nonumber
\end{eqnarray}
The corresponding correction for the integrated spectral density $N^V(\Lambda_1,\Lambda_2,m)_{NLO}$ for $\Lambda_1=m$ is given by 
\begin{eqnarray}
\label{eq:N_FSE}
\Delta N^V(m,\Lambda_2,m)_{NLO} & = & \int_{m}^{\Lambda_2}[\Delta\rho_Q^V(\lambda,m)+\Delta\rho_Q^V(-\lambda,m)]_{NLO}d\lambda=
\frac{2\Sigma^2}{\pi}\frac{\sqrt{\Lambda_2^2-m^2}}{(4\pi)^2F^4}   \\
&& \sum_{\{n1,n2,n3,n4\}\neq 0}\Bigg\{\frac{2}{\sqrt{\Lambda_2^2-m^2}}{\rm Im}\left[F_{-2}\left(\frac{\Sigma q_n^2}{4F^2},i\sqrt{\Lambda_2^2-m^2}+m\right)    \right]\nonumber\\
& - & \frac{m}{\sqrt{\Lambda_2^2-m^2}}{\rm Im}\left[F_{-1}\left(\frac{\Sigma q_n^2}{2F^2},i\sqrt{\Lambda_2^2-m^2}\right)\right] \nonumber\\
&+ &{\rm Re}\left[F_{-1}\left(\frac{\Sigma q_n^2}{2F^2},i\sqrt{\Lambda_2^2-m^2}\right)\right]\Bigg\}.\nonumber
\end{eqnarray}

\end{appendix}

\bibliographystyle{apsrev}
\bibliography{biblio}

\end{document}